\documentclass[Physsubmission, Phys]{SciPost}
\hypersetup{
    colorlinks,
    linkcolor={red!50!black},
    citecolor={blue!50!black},
    urlcolor={blue!80!black}
}

\usepackage[bitstream-charter]{mathdesign}
\DeclareSymbolFont{usualmathcal}{OMS}{cmsy}{m}{n}
\DeclareSymbolFontAlphabet{\mathcal}{usualmathcal}

\begin{document}

% TODO: write your article's title here.
% The article title is centered, Large boldface, and should fit in two lines
\begin{center}{\Large \textbf{
Masses of vector and pseudovector hybrid mesons in a chiral symmetric model\\
}}\end{center}

% TODO: write the author list here. Use initials + surname format.
% Separate subsequent authors by a comma, omit comma at the end of the list.
% Mark the corresponding author with a superscript *.
\begin{center}
Walaa I. Eshraim\textsuperscript{$\star$}
\end{center}

% TODO: write all affiliations here.
% Format: institute, city, country
\begin{center}
 New York University Abu Dhabi, Saadiyat Island, P.O. Box 129188, Abu Dhabi, U.A.E\\
% TODO: provide email address of corresponding author
* wie2003@nyu.edu
\end{center}

\begin{center}
\today
\end{center}

% For convenience during refereeing (optional),
% you can turn on line numbers by uncommenting the next line:
%\linenumbers
% You should run LaTeX twice in order for the line numbers to appear.

\definecolor{palegray}{gray}{0.95}
\begin{center}
\colorbox{palegray}{
  \begin{minipage}{0.95\textwidth}
    \begin{center}
    {\it  XXXIII International (ONLINE) Workshop on High Energy Physics \\“Hard Problems of Hadron Physics:  Non-Perturbative QCD \& Related Quests”}\\
    {\it November 8-12, 2021} \\
    \doi{10.21468/SciPostPhysProc.?}\\
    \end{center}
  \end{minipage}
%\end{tabular}
}
\end{center}

\section*{Abstract}
{\bf
% TODO: write your abstract here.
We enlarge the chiral model, the so-called extended Linear Sigma Model (eLSM), by including the low-lying hybrid nonet with exotic quantum numbers $J^{PC}=1^{-+}$ and the nonet of their chiral partners with $J^{PC}=1^{+-}$ to a global 
$U(3)_r \times U(3)_l$ chiral symmetry. We use the assignment of the $\pi_1^{hyb}= \pi_1(1600)$ as input to determine the unknown parameters. Then, we compute the lightest vector and pseudovector hybrid masses that could guide ongoing and upcoming experiments in searching for hybrids.
}

% TODO: include a table of contents (optional)
% Guideline: if your paper is longer that 6 pages, include a TOC
% To remove the TOC, simply cut the following block
%\vspace{10pt}
%\noindent\rule{\textwidth}{1pt}
%\tableofcontents\thispagestyle{fancy}
%\noindent\rule{\textwidth}{1pt}
%\vspace{10pt}

\section{Introduction}
\label{sec:intro}
% TODO: write your article here.
The investigation of the properties of exotic quarkonia, the so-called hybrids, is extremely interesting and an important step toward the understanding of the nontraditional hadronic states, i.e., those structures beyond the normal meson and baryon, which are allowed in the framework of quantum chromodynamics (QCD) \cite{Gross, Politzer, Wilson} and quark model \cite{Gell-Mann, Zweig}. Hybrids are colour singlets and constitute of quark-antiquark pair and gluonic degree of freedom. In Lattice QCD, a rich spectrum of hybrid states are predicted below 5 GeV \cite{Dudek, Dudek2, Dudek3}, but there are still no predominantly hybrid states assigned to be one of the
listed mesons in the PDG \cite{pdg}. Quite interestingly, recent results by
COMPASS concerning the confirmation of the state $\pi_{1}(1600)$ with exotic
quantum numbers $1^{-+}$ led to a revival of interest in this topic \cite{COMPASS}.

In this work, we investigate vector hybrids by enlarging the extended Linear
Sigma Model (eLSM) \cite{dick}. In particular, we make predictions
for a nonet of exotic hybrids with quantum numbers $J^{PC}=1^{-+}$. Moreover,
we also make predictions for the nonet of their chiral partners, with quantum
numbers $J^{PC}=1^{+-}.$

The eLSM has shown to be able to describe various hadronic masses and decays
below 1.8 GeV, as the fit in Ref. \cite{dick} confirms, hence it represents a
solid basis to investigate states that go beyond the simple $\bar{q}q$
picture. In the past, various non-conventional mesons were studied in the
eLSM. Namely, the scalar glueball is automatically present in the eLSM as a
dilaton and is coupled to light mesons: it represents an important element of
the model due to the requirement of dilatation invariance (as well as its
anomalous breaking)\cite{staninew}. The eLSM has been used to study the pseudoscalar
glueball \cite{walaaG1, walaaG11, walaaG4, walaaG5, walaaG6}, the first excited pseudoscalar glueball \cite{walaaG2, walaaG3}, and hybrids \cite{walaah}. Moreover, the connection and compatibility with chiral perturbation theory
\cite{Divotgey}, as well as the extention to charmed mesons
\cite{walaac, walaac1, walaac1a, walaac1b, walaac1b1, walaac2, walaac2a, walaac2b} and the inclusion of baryons in the so-called mirror assignment
\cite{gallas, olbrich} were performed.

In the present study, we extend the eLSM to hybrids by constructing the chiral multiplet for hybrid nonets with $J^{PC}=1^{-+}$ and $J^{PC}=1^{+-}$ and determine the interaction terms which satisfy chiral symmetry. Consequently, the spontaneous symmetry breaking is responsible for mass differences between the $1^{+-}$ crypto-exotic hybrids and the lower-lying $1^{-+}$. We work out the masses of vector and pseudovector hybrid mesons.

\section{Hybrid mesons in the chiral model}
In this section, we enlarge the eLSM Lagrangian by including hybrid mesons in the case of $N_f=3$ 

\begin{equation}
\mathcal{L}_{eLSM}^{\text{with hybrids}}=\mathcal{L}_{eLSM}+\mathcal{L}%
_{eLSM}^{\text{ hybrid}}%
\end{equation}
where $\mathcal{L}_{eLSM}$ is the standard of the eLSM Lagrangain, which are constructed under chiral and
dilatation symmetries, as well as their explicit and spontaneous breaking
features (for more details see Refs.\cite{dick}).\\
We introduce the hybrids in the eLSM as:%

\begin{align}
\mathcal{L}_{eLSM}^{\text{ hybrid}}&=\mathcal{L}%
_{eLSM}^{\text{ hybrid-quadratic}}+\mathcal{L}_{eLSM}^{\text{ hybrid-linear}}\,\nonumber\\
&=\mathcal{L}%
_{eLSM}^{\text{ hybrid-kin}}+\mathcal{L}_{eLSM}^{\text{ hybrid-mass}}+\mathcal{L}_{eLSM}^{\text{ hybrid-linear}}\,
\end{align}
where the $\mathcal{L}%
_{eLSM}^{\text{ hybrid-kin}}$ and $\mathcal{L}_{eLSM}^{\text{ hybrid-linear}}$ terms are described in details in Ref.\cite{walaah}. The masses of hybrids can be extracted from the following mass term

\begin{align}
 \mathcal{L}_{eLSM}^{\text{ hybrid-mass}}=&\,m_{1}^{hyb,2}\frac{G^{2}%
}{G_{0}^{2}}\mathrm{Tr}\left(  L_{\mu}^{hyb,2}+R_{\mu}^{hyb,2}\right)
+\mathrm{Tr}\left(  \Delta^{hyb}\left(  L_{\mu}^{hyb,2}+R_{\mu}^{hyb,2}%
\right)  \right) \nonumber\\
& +\, \frac{h_{1}^{hyb}}{2}\mathrm{Tr}(\Phi^{\dagger}\Phi)\mathrm{Tr}\left(
L_{\mu}^{hyb,2}+R_{\mu}^{hyb,2}\right)  +h_{2}^{hyb}\mathrm{Tr}[\left\vert
L_{\mu}^{hyb}\Phi\right\vert ^{2}+\left\vert \Phi R_{\mu}^{hyb}\right\vert
^{2}]\nonumber \\
&+\, 2h_{3}^{nyb}\mathrm{Tr}(L_{\mu}^{hyb}\Phi R^{hyb,\mu}\Phi^{\dagger
})\text{ .}\label{Lag}%
\end{align}
which satisfies both chiral and dilatation invariance. $G$ is the dilaton field and $G_0$ its vacuum's expectation value. The multiplet of the scalar and pseudoscalar mesons, $\Phi$, is defined as

\begin{equation}
\Phi=S+i P=\frac{1}{\sqrt{2}}\left(
\begin{array}
[c]{ccc}%
\frac{\sigma_{N}+a_{0}^{0}}{\sqrt{2}} & a_{0}^{+} & K_{S}^{+}\\
a_{0}^{-} & \frac{\sigma_{N}-a_{0}^{0}}{\sqrt{2}} & K_{S}^{0}\\
K_{S}^{-} & \bar{K}_{S}^{0} & \sigma_{S}%
\end{array}
\right) +i \frac{1}{\sqrt{2}}\left(
\begin{array}
[c]{ccc}%
\frac{\eta_{N}+\pi^{0}}{\sqrt{2}} & \pi^{+} & K^{+}\\
\pi^{-} & \frac{\eta_{N}-\pi^{0}}{\sqrt{2}} & K^{0}\\
K^{-} & \bar{K}^{0} & \eta_{S}%
\end{array}
\right) \text{ ,}%
\end{equation}
and transforms under chiral transformations $U_{L}(3)\times
U_{R}(3)$: $\Phi\rightarrow U_{L}\Phi U_{R}^{\dagger}$, where $U_{L}$ and
$U_{R}$ are $U(3)$, under parity $\Phi\rightarrow\Phi^{\dagger}$ and
under charge conjugation $\Phi\rightarrow\Phi^{t}.$ \\
(i) The scalar fields are $\{a_{0}%
(1450),K_{0}^{\ast}(1430),\sigma_{N},\sigma_{S}\}$ with quantum number $J^{PC}=0^{++}$ \cite{pdg}, and lie above $1$ GeV \cite{dick}, where the non-strange bare field $\sigma_{N}%
\equiv\left\vert \bar{u}u+\bar{d}d\right\rangle /\sqrt{2}$ corresponds
predominantly to the resonance \thinspace$f_{0}(1370)$ and the bare field
$\sigma_{S}\equiv\left\vert \bar{s}s\right\rangle $ predominantly to
$f_{0}(1500).$ Finally, in the eLSM the state $f_{0}(1710)$ is predominantly a
scalar glueball, see details in\ Ref. \cite{staninew}. 
(ii) The pseudoscalar fields are $\{\pi$,
$K,\eta,\eta^{\prime}\}$ with quantum numbers $J^{PC}=0^{-+}$ \cite{pdg}, where $\eta$ and $\eta^{\prime}$ arise
via the mixing $\eta=\eta_{N}\cos\theta_{p}+\eta_{S}\sin\theta_{p},$
$\eta^{\prime}=-\eta_{N}\sin\theta_{p}+\eta_{S}\cos\theta_{p}$ with
$\theta_{p}\simeq-44.6^{\circ}$ \cite{dick}.\\
We now turn to the right-handed and left-handed, $R_{\mu}^{hyb} \text{and}\, L_{\mu}^{hyb}$, combinations of exotic hybrid states, which combine the vector fields in the hybrid sector $\Pi_{ij}^{hyb,\mu}$ with the pseudovector fileds in the hybrid sector $B_{ij}^{hyb,\,\mu}$.\\
The hybrid sector $\Pi_{ij}^{hyb,\mu}$ is vector currents with one additional gluon with quantum numbers $J^{PC}=1^{-+}$, and is given by 
\begin{equation}
\Pi_{ij}^{hyb,\mu}=\frac{1}{\sqrt{2}}\bar{q}_{j}G^{\mu\nu}\gamma_{\nu}%
q_{i}=\Pi^{hyb,\mu}=\frac{1}{\sqrt{2}}\left(
\begin{array}
[c]{ccc}%
\frac{\eta^{hyb}_{1,N}+\pi_{1}^{hyb,0}}{\sqrt{2}} & \pi_{1}^{hyb+} & K_{1}^{hyb+}\\
\pi_{1}^{hyb-} & \frac{\eta_{1,N}^{hyb}+\pi_{1}^{hyb,0}}{\sqrt{2}} & K_{1}^{hyb,0}\\
K_{1}^{hyb,-} & \bar{K}_{1}^{hyb,0} & \eta_{1,S}^{hyb}%
\end{array}
\right)  ^{\mu}\;\text{, }
\end{equation}
where the gluonic field tensor $G^{\mu\nu}$ is equal to $\partial^{\mu}A^{\nu}-\partial^{\mu}A^{\nu}-g_{QCD}[A^{\mu
},A^{\nu}]$, and $\Pi^{hyb,\mu}$ contains $\{\pi(1600), \, K_1(?),\, \eta_1(?),\,\eta_1(?)\}$ which only the isovector member corresponds to a physical resonance at the present. The exotic hybrid field $\pi_{1}$ is assigned
to $\pi_{1}(1600),$ (the details of this assignment are given in Ref. \cite{JPAC}).
There are not yet candidates for the other members of the nonet, but we shall estimate their masses in Sec. 3.\\
The pseudovector fields, $B_{ij}^{hyb,\mu}$ in the hybrid sector, after including the gluon field, with quantum numbers $J^{PC}=1^{+-}$, is written  as
\begin{equation}
B_{ij}^{hyb,\mu}=\frac{1}{\sqrt{2}}\bar{q}_{j}G^{\mu\nu}\gamma^{5}\gamma_{\nu
}q_{i}=B^{hyb,\mu}=\frac{1}{\sqrt{2}}\left(
\begin{array}
[c]{ccc}%
\frac{h_{1N,B}^{hyb}+b_{1}^{hyb,0}}{\sqrt{2}} & b_{1}^{hyb,+} & K_{1,B}%
^{hyb+}\\
b_{1}^{hyb,+} & \frac{h_{1N,B}^{hyb}-b_{1}^{hyb,0}}{\sqrt{2}} & K_{1,B}%
^{hyb0}\\
K_{1,B}^{hyb-} & \bar{K}_{1,B}^{hyb0} & h_{1S,B}^{hyb}%
\end{array}
\right)  ^{\mu} \text{ .}%
\end{equation}
The nonet $B_{ij}^{hyb,\mu}$ has not yet any experimental candidate. So, all fields $\{b_1(?),\,K_{1,B}(?),\, h_1(?),\, h_1(?)\}$ are unkown yet. In the lattice calculation of  Ref. \cite{Dudek2}, an upper limit of about $2.4$ GeV is reported,
but lattice simulation still used a quite large pion mass.  We estimate the
mass of the $b_{1}^{hyb}$ state, the chiral partner of $\pi_{1},$ to a value of about (or eventually somewhat larger than) $2$ GeV. For definiteness, we shall assign it to an hypothetical state $b_{1}(2000?)$
state. The other member masses of the pseudovector crypto-exotic nonet follow as a consequence of this assumption.
One can obtain the right-handed and left-handed currents as follows
$$R_{\mu}^{hyb}=\Pi^{hyb,\mu}-B_{ij}^{hyb,\mu} \text{and}\, L_{\mu}^{hyb}=\Pi^{hyb,\mu}+B_{ij}^{hyb,\mu}$$
and transform as $R_{\mu}^{hyb}\rightarrow U_R R_{\mu}^{hyb} U_R^\dagger$ and $L_{\mu}^{hyb}\rightarrow U_L L_{\mu}^{hyb} U_L^\dagger$ and under parity as $R_{\mu}^{hyb}\rightarrow L^{\mu,\,hyb}$ and $L_{\mu}^{hyb}\rightarrow R^{\mu,\,hyb}$ as well as under C as $R_{\mu}^{hyb}\rightarrow L^{hyb,\mu,t}$ and $L_{\mu}^{hyb}\rightarrow R^{hyb,\mu,t}$.
See Ref. \cite{walaah} for more details and discussions.\\

\section{Masses of hybrids}

Masses of hybrids can be calculated from the expression (\ref{Lag}) by taking into account that the multiplet of the scalar and pseudoscalar fields, $\Phi$, has a nonzero condensate or vacuum's expectation value. Consequently, the spontaneous symmetry breaking is reflected from that condensate. Especially relevant is
the term $h_{3}^{nyb}$ which generates a mass difference between the $1^{-+}$
and $1^{+-}$ hybrids, after shifting the latter masses upwards (see Ref. \cite{walaah}). Note, the second term breaks explicitly flavor symmetry (direct contribution
to the masses due to nonzero bare quark masses):
\begin{equation}
\Delta^{hyb}=diag\{\delta_{N}^{hyb},\delta_{N}^{hyb},\delta_{S}^{hyb}%
\}\text{.}%
\end{equation}

After a straightforward calculation, the (squared) masses of the $1^{-+}$
exotic hybrid mesons and the (squared) masses of the cryptoexotic pseudovector hybrid states were obtained as seen in Ref. \cite{walaah}. Consequently, one can get the (exact) relations as
\begin{align}
m_{b_{1}^{hyb}}^{2}-m_{\pi_{1}}^{2}  &  =-2h_{3}^{hyb}\phi_{N}^{2}%
\label{hcp1}\\
m_{K_{1,B}^{hyb}}^{2}-m_{K_{1}}^{2}  &  =-\sqrt{2}\phi_{N}\phi_{S}h_{3}%
^{hyb}\label{hcp2}\\
m_{h_{1S}^{hyb}}^{2}-m_{\eta_{1,S}}^{2}  &  =-h_{3}^{hyb}\phi_{S}^{2}
\label{hcp3}%
\end{align}
As seen in Eqs. (\ref{hcp1}-\ref{hcp3}), the parameter $h_{3}^{hyb}$ is the only parameter responsible for the mass splitting of the hybrid chiral partners. After fixing all the parameters that appear in the Lagrangian (\ref{Lag}) and the square masses equations (see details in Ref. \cite{walaah}), we obtain the following results ( shown in Table 1) for the masses of the vector and pseudovector hybrid mesons:

\begin{table}[h] \centering
%EndExpansion%
\begin{tabular}
[c]{|c|c|}\hline
Resonance &$ Mass [MeV] $\\\hline
$\Pi_1^{hyb}$ & $1600$ [input using $\pi_1(1600)$] \cite{} \\\hline
$\eta_{1,N}^{hyb}$ & $1660$ \\\hline
$\eta_{1,S}^{hyb}$ & $1751$  \\\hline
$K_1^{hyb}$ & $1707$  \\\hline
$b_1^{hyb}$ & $2000$ [input set as an estimate] \\\hline
$h_{1N,B}^{hyb}$ & $2000$  \\\hline
$K^{hyb}_{1,B}$ & $2063$  \\\hline
$h_{1S,B}^{hyb}$ & $2126$  \\\hline
\end{tabular}%
%TCIMACRO{\TeXButton{Caption}{\caption
%{Summary of the quark-antiquark and hybrid nonets and their properties.}}}%
%BeginExpansion
\caption
{Masses of the exotic $J^{PC}=1^{-+}$ and $J^{PC}=1^{+-}$ hybrid mesons.}%
%EndExpansion%
%TCIMACRO{\TeXButton{E}{\end{table}}}%
%BeginExpansion
\end{table}%

\section{Conclusion}

We have enlarged a chiral model, the so-called eLSM, in the case of $N_f=3$ by including the hybrid state, the lightest hybrid nonet with $J^{PC}=1^{-+}$ and of its chiral partner with $J^{PC}=1^{+-}$, into a chiral multiplet. The eLSM implements the global chiral $U(N_f)_r \times U(N_f)_l$ symmetry and the symmetries of QCD: the discrete T, P, and C symmetries. The global chiral symmetry is broken in several ways: explicitly through non-vanishing quark masses, spontaneously due to the chiral condensate, and at the quantum level due to the chiral anomaly. To our knowledge, this is the first time that a model was constructed, which contains vector and pseudovector hybrid mesons. The resonance $\pi^{hyb}_1$ is assigned to $\pi_1(1600)$ (with mass $1660^{+15}_{-11}$ MeV) and $b_1^{hyb}$ is set to $2$ GeV. The masses of the other hybrid states are computed and their results are reported in Table $1$. Note that our model predicts the mass of the state $\eta_1^{hyb}$ to be the same as $\pi_1^{hyb}\equiv \pi(1600)$ because of the small mixing of the nonstrange-strange quarks, which is in agreement with the homochiral nature of the chiral multiplet. 
Moreover, the calculation and the results of the decay widths of the lightest vector and pseudovector hybrid mesons are presented in
Ref. \cite{walaah}.

\section*{Acknowledgements}
The author thanks F. Giacosa, C. Fischer, and D. Parganlija for cooperation leading to Ref. \cite{walaah}. Moreover, the author acknowledges support from the NYUAD Center for Interacting Urban Networks (CITIES) through Tamkeen under the NYUAD Research Institute Award CG001.


\begin{thebibliography}{99}                                                                                              

%\cite{Gross}
\bibitem{Gross}
D.~J.~Gross and F.~Wilczek,
%``Ultraviolet Behavior of Nonabelian Gauge Theories,''
Phys. Rev. Lett. \textbf{30}, 1343-1346 (1973).

\bibitem{Politzer}
H.~D.~Politzer,
%``Reliable Perturbative Results for Strong Interactions?,''
Phys. Rev. Lett. \textbf{30}, 1346-1349 (1973).

\bibitem{Wilson}
K.~G.~Wilson,
%``Confinement of Quarks,''
Phys. Rev. D \textbf{10}, 2445-2459 (1974).

\bibitem{Gell-Mann}
M.~Gell-Mann,
%``A Schematic Model of Baryons and Mesons,''
Phys. Lett. \textbf{8}, 214-215 (1964).

\bibitem{Zweig} G. Zweig, Report No. CERN-TH-401.

%%%%%%%%%%%%%%%%%%%%%%%%%%%%%%%%%%%%%%%%
\bibitem{Dudek}
J.~J.~Dudek, R.~G.~Edwards, M.~J.~Peardon, D.~G.~Richards and C.~E.~Thomas,
%``Highly excited and exotic meson spectrum from dynamical lattice QCD,''
Phys. Rev. Lett. \textbf{103} (2009), 262001
%doi:10.1103/PhysRevLett.103.262001
[arXiv:0909.0200 [hep-ph]].

\bibitem{Dudek2}
J.~J.~Dudek, R.~G.~Edwards, M.~J.~Peardon, D.~G.~Richards and C.~E.~Thomas,
%``Toward the excited meson spectrum of dynamical QCD,''
Phys. Rev. D \textbf{82} (2010), 034508
%doi:10.1103/PhysRevD.82.034508
[arXiv:1004.4930 [hep-ph]].

\bibitem{Dudek3}
J.~J.~Dudek \textit{et al.} [Hadron Spectrum],
%``Toward the excited isoscalar meson spectrum from lattice QCD,''
Phys. Rev. D \textbf{88} (2013) no.9, 094505
%doi:10.1103/PhysRevD.88.094505
[arXiv:1309.2608 [hep-lat]].

%%%%%%%%%%%%%%%%%%%%%%%%%%%%%%%%%%%%%%55

\bibitem {pdg}C. Patrignani et al. (Particle Data group), Chin. Phys. C, 40,
100001 (2016).

\bibitem{COMPASS}
M.~Aghasyan \textit{et al.} [COMPASS],
%``Light isovector resonances in $\pi^- p \to \pi^-\pi^-\pi^+ p$ at 190 GeV/${\it c}$,''
Phys. Rev. D \textbf{98} (2018) no.9, 092003
%doi:10.1103/PhysRevD.98.092003
[arXiv:1802.05913 [hep-ex]].

\bibitem {dick}
%\cite{Parganlija:2012fy}
%\bibitem{Parganlija:2012fy}
D.~Parganlija, P.~Kovacs, G.~Wolf, F.~Giacosa and D.~H.~Rischke,
%``Meson vacuum phenomenology in a three-flavor linear sigma model with (axial-)vector mesons,''
Phys.\ Rev.\ D \textbf{87} (2013) 014011 [arXiv:1208.0585 [hep-ph]].
%%CITATION = ARXIV:1208.0585;%%

\bibitem {staninew}
S.~Janowski, F.~Giacosa and D.~H.~Rischke,
%``Is f0(1710) a glueball?,''
Phys. Rev. D \textbf{90} (2014) no.11, 114005
%doi:10.1103/PhysRevD.90.114005
[arXiv:1408.4921 [hep-ph]].

%%%%%%%%%%%%%%%%%%%%%%%%%%%%%%%%%5

\bibitem {walaaG1}
%\cite{Eshraim:2012jv}
%\bibitem{Eshraim:2012jv}
W.~I.~Eshraim, S.~Janowski, F.~Giacosa and D.~H.~Rischke,
%``Decay of the pseudoscalar glueball into scalar and pseudoscalar mesons,''
Phys.\ Rev.\ D \textbf{87} (2013) 5, 054036 [arXiv:1208.6474 [hep-ph]].
%%CITATION = ARXIV:1208.6474;%%

\bibitem {walaaG11}
%\cite{Eshraim:2012rb}
%\bibitem{Eshraim:2012rb}
W.~I.~Eshraim, S.~Janowski, A.~Peters, K.~Neuschwander and F.~Giacosa,
%``Interaction of the pseudoscalar glueball with (pseudo)scalar mesons and nucleons,''
Acta Phys.\ Polon.\ Supp.\ \textbf{5} (2012) 1101 [arXiv:1209.3976 [hep-ph]].
%%CITATION = ARXIV:1209.3976;%%

\bibitem{walaaG4}
W.~I.~Eshraim,
%``Decay of the pseudoscalar glueball into vector, axial-vector, scalar and pseudoscalar mesons,''
[arXiv:2005.11321 [hep-ph]].


\bibitem{walaaG5}
W.~I.~Eshraim,
%``Interaction of the pseudoscalar glueball with (pseudo)scalar mesons and their first excited states,''
PoS \textbf{PANIC2021} (2022), 173.
%doi:10.22323/1.380.0173
%0 citations counted in INSPIRE as of 15 Apr 2022


\bibitem{walaaG6}
W.~I.~Eshraim,
%``A pseudoscalar glueball and charmed mesons in the extended Linear Sigma Model,''
EPJ Web Conf. \textbf{95} (2015), 04018
%doi:10.1051/epjconf/20159504018
[arXiv:1411.2218 [hep-ph]].

\bibitem {walaaG2}
%\cite{Eshraim:2016mds}
%\bibitem{Eshraim:2016mds}
W.~I.~Eshraim and S.~Schramm,
%``Decay modes of the excited pseudoscalar glueball,''
Phys.\ Rev.\ D \textbf{95} (2017) no.1, 014028
%doi:10.1103/PhysRevD.95.014028
[arXiv:1606.02207 [hep-ph]].

\bibitem{walaaG3}
W.~I.~Eshraim,
%``Decay of the pseudoscalar glueball and its first excited state into scalar and pseudoscalar mesons and their first excited states,''
Phys. Rev. D \textbf{100} (2019) no.9, 096007
%doi:10.1103/PhysRevD.100.096007
[arXiv:1902.11148 [hep-ph]].

%%%%%%%%%%%%%%%%%%%%%%%%%55555
\bibitem{walaah}
W.~I.~Eshraim, C.~S.~Fischer, F.~Giacosa and D.~Parganlija,
%``Hybrid phenomenology in a chiral approach,''
Eur. Phys. J. Plus \textbf{135} (2020) no.12, 945
%doi:10.1140/epjp/s13360-020-00900-z
[arXiv:2001.06106 [hep-ph]].
%%%%%%%%%%%%%%%%%%%%%%%%%%%

\bibitem{Divotgey}
F.~Divotgey, P.~Kovacs, F.~Giacosa and D.~H.~Rischke,
%``Low-energy limit of the extended Linear Sigma Model,''
Eur. Phys. J. A \textbf{54} (2018) no.1, 5
%doi:10.1140/epja/i2018-12458-9
[arXiv:1605.05154 [hep-ph]].


%%%%%%%%%%%%%%%%%%%%%%%%%%%%%%%%%5555
\bibitem {walaac}
 W.~I.~Eshraim,
%``Masses of light and heavy mesons in a $U(4)_r \times U(4)_l$ linear sigma model,''
PoS QCD {\bf -TNT-III} (2013) 049
[arXiv:1401.3260 [hep-ph]].

\bibitem {walaac1}
%\cite{Eshraim:2014eka}
%\bibitem{Eshraim:2014eka}
W.~I.~Eshraim, F.~Giacosa and D.~H.~Rischke,
%``Phenomenology of charmed mesons in the extended Linear Sigma Model,''
Eur.\ Phys.\ J.\ A \textbf{51} (2015) 9, 112 [arXiv:1405.5861 [hep-ph]].

\bibitem {walaac1a}
 W.~I.~Eshraim,
%``Vacuum properties of open charmed mesons in a chiral symmetric model,''
J.\ Phys.\ Conf.\ Ser.\  {\bf 599} (2015) no.1,  012009
[arXiv:1411.4749 [hep-ph]].

\bibitem {walaac1b}
 W.~I.~Eshraim and F.~Giacosa,
%``Decays of open charmed mesons in the extended Linear Sigma Model,''
EPJ Web Conf.\  {\bf 81} (2014) 05009
[arXiv:1409.5082 [hep-ph]].

\bibitem{walaac1b1}
W.~I.~Eshraim,
%``Vacuum properties of open charmed mesons in a chiral symmetric model,''
J. Phys. Conf. Ser. \textbf{599} (2015) no.1, 012009
%doi:10.1088/1742-6596/599/1/012009
[arXiv:1411.4749 [hep-ph]].

\bibitem {walaac2}
 W.~I.~Eshraim 
  %``Decay of charmonium states into a scalar and a pseudoscalar glueball,''
  EPJ Web Conf.\  {\bf 126} (2016) 04017.
  %%CITATION = doi:10.1051/epjconf/201612604017;%%
  %1 citations counted in INSPIRE as of 27 Jul 2018
  .
\bibitem {walaac2a}
 W.~I.~Eshraim and C.~S.~Fischer,
  %``Hadronic decays of the (pseudo-)scalar charmonium states $\eta_{c}$ and $\chi_{c0}$ in the extended Linear Sigma Model,''
  Eur.\ Phys.\ J.\ A {\bf 54} (2018) no.8,  139
  [arXiv:1802.05855 [hep-ph]].
  %%CITATION = ARXIV:1802.05855;%%

\bibitem{walaac2b}
W.~I.~Eshraim,
%``Decay of the scalar charmonium state $\chi_{c0}(IP)$ in the extended Linear Sigma Model,''
PoS \textbf{CHARM2020} (2021), 056.

%%%%%%%%%%%%%%%%%%%%%%%%%%%%%%%%%%%%%%%%%%%%%%%%%%55
\bibitem {gallas}S.~Gallas, F.~Giacosa and D.~H.~Rischke,
%``Vacuum phenomenology of the chiral partner of the nucleon in a linear sigma
%model with vector mesons,''
Phys.\ Rev.\ D \textbf{82} (2010) 014004 [arXiv:0907.5084 [hep-ph]].
%\cite{Olbrich:2015gln}
\bibitem{olbrich}
L.~Olbrich, M.~Z\'{e}t\'{e}nyi, F.~Giacosa and D.~H.~Rischke,
%``Three-flavor chiral effective model with four baryonic multiplets within the mirror assignment,''
Phys.\ Rev.\ D \textbf{93} (2016) no.3, 034021
%doi:10.1103/PhysRevD.93.034021
[arXiv:1511.05035 [hep-ph]].
%%CITATION = doi:10.1103/PhysRevD.93.034021;%%
%\cite{Gallas:2013ipa}
%\bibitem{Gallas:2013ipa}

%%%%%%%%%%%%%%%%%%%%%%%%%%%%%%%5

\bibitem{JPAC}
A.~Rodas \textit{et al.} [JPAC],
%``Determination of the pole position of the lightest hybrid meson candidate,''
Phys. Rev. Lett. \textbf{122}, no.4, 042002 (2019)
%doi:10.1103/PhysRevLett.122.042002
[arXiv:1810.04171 [hep-ph]].



\end{thebibliography}
\end{document}